# Analytical and Learning-Based Spectrum Sensing

# Time Optimization in Cognitive Radio Systems

Hossein Shokri-Ghadikolaei[†], Younes Abdi, Masoumeh Nasiri-Kenari[†]

[†]Wireless Research Lab., Elec. Eng. Dept., Sharif University of Technology

## Abstract

Powerful spectrum sensing schemes enable cognitive radios (CRs) to find transmission opportunities in spectral resources allocated exclusively to the primary users. In this paper, maximizing the average throughput of a secondary user by optimizing its spectrum sensing time is formulated assuming that a prior knowledge of the presence and absence probabilities of the primary users is available. The energy consumed for finding a transmission opportunity is evaluated and a discussion on the impact of the number of the primary users on the secondary user throughput and consumed energy is presented. In order to avoid the challenges associated with the analytical method, as a second solution, a systematic neural network-based sensing time optimization approach is also proposed. The proposed adaptive scheme is able to find the optimum value of the channel sensing time without any prior knowledge or assumption about the wireless environment including the primary users' presence probabilities. The structure, performance, and cooperation of the artificial neural networks used in the proposed method are disclosed in detail and a set of illustrative simulation results is presented to validate the analytical results as well as the performance of the proposed learning-based optimization scheme.

put.

# I. INTRODUCTION

Opportunistic spectrum sharing has been developed through the promising concept of cognitive
radio (CR) to meet the ever-growing spectrum demand for new wireless services. Conceptually,
CR is an adaptive communication system which offers the promise of intelligent radios that
can learn from and adapt to their environments [1]. The major issue in designing a cognitive
radio network is to protect incumbent/primary users from potential interference problems while
providing acceptable quality-of-service (QoS) levels for secondary users (SUs) (i.e., unlicensed
users). To this end, sensing capability is exploited in CRs which enable them to find some
transmission opportunities called white spaces. These white spaces are temporarily-available
spectrums which are not used by the primary users (PUs). In fact, incorporating appropriate
spectrum sensing schemes enable CRs to transmit and/or receive data while no channels are
dedicated to them. Average throughput of the CRs, average sensing time, and consumed energy
are some common metrics considered in designing effective sensing schemes.

Throughput maximization through optimizing spectrum sensing time has gained a lot of
interest. Spectrum sensing time is one of the most effective factors which must be determined
carefully, to obtain a powerful sensing scheme. In [2], [3], [4], and [5], the impact of spectrum
sensing time on the overall throughput of the SUs is investigated. It is shown in [2] that as the
sensing time increases, the sensing accuracy increases as well, but the throughput decreases; thus
there is a tradeoff. In [3], the optimum value of the sensing time has been found numerically,

while the spectrum mobility effect has not been taken into account. Spectrum mobility introduces a new concept of handoff that is called Spectrum Handover, or simply hand-over (HO). In this case, the SU must leave the spectrum and continue its transmission on another spectrum when a PU arrives. Due to interruptions from the PUs, it is likely to have multiple spectrum HOs within the transmission period of a secondary connection. Clearly, multiple HOs will increase the overall sensing time [4]. In [5], it is assumed that the licensed spectrums are numbered sequentially and the SU starts to sense the spectrums from the top of a list, and in the case of occupation, the SU senses the next one and this process is continued until an idle spectrum is found. Based on this assumption, an optimization problem is formulated in [5] in order to minimize the average sensing time. Although the false detection and spectrum handover effects on sensing time have been investigated in [5], the adverse effect of the handover (the sensing time effect) on the SU throughput has not been taken into account.

In this paper, two independent solution procedures are proposed for this problem. The first solution is the conventional analytical procedure which is obtained through the mathematical formulation of the SU's behavior. To this end, the exact mathematical evaluation of the average throughput of the SU is presented in this paper. Then, a novel optimization problem is derived to find the optimum sensing time for maximizing the SU throughput. The tradeoff between the maximum achievable throughput and the sensing energy consumption is explained and a design parameter is introduced to modify the optimization problem in order to address the consumed energy.

It is worth noting that the optimum value of the sensing time derived through conventional analytical optimization procedures depends directly on the models adopted for the channel, the users' traffic, and the PUs' behavior. Despite of the increased analytical complexity introduced by using more complete modelings, these models are not necessarily consistent with the actual environments in which the CRs work, and therefore, the derived values cannot be considered as the perfect optimum ones. Moreover, if any changes occur in the parameters describing the wireless environment and traffic conditions experienced by the SUs, it is required to estimate the new parameters, repeat the analysis, and recalculate the optimum values of the sensing time. In order to deal with these issues appropriately, the second method is proposed which is based on systematic configuration of two kinds of well-known artificial neural networks. Specifically, a Multilayer Feed-Forward (MFF) neural network [6] is used to replace the mathematical modelling, by learning the actual behavior of the secondary link, i.e., the effect of the spectrum sensing time on the average throughput of the SU. Based on the actual (non-analytical) model of the link which is learned by the MFF network, a Kennedy-Chua (KC) neural network [7] is used to find the optimum value of the spectrum sensing time. This learning-based optimization scheme does have several advantages over the analytical method. First, no prior knowledge about the link behavior, such as the presence or absence probabilities of primary users are required. Second, the limited consistency of the mathematical models with the real wireless environment does not affect the optimality of the derived spectrum sensing time. Third, using this learning-based optimization scheme, an adaptive system is proposed which is capable of effectively following the variations in the link and keeps the average throughput at the maximum level in non-stationary conditions.

The rest of this paper is organized as follows. In Section II, we describe the considered CR system model and derive the related optimization problem. Besides, the tradeoff between the SU throughput and its consumed energy is investigated. In Section III, the utilized neural networks as well as their structures and performance are introduced in detail. Moreover, the proposed learning-based sensing time optimization scheme is described and its operation algorithm is presented. Numerical results are then presented in Section IV, followed by concluding remarks provided in Section V.

## II. System Model and The Related Optimization Problem

### A. System Model

We assume a primary network with $N_p$ users, each of them with a dedicated channel. Therefore, there are $N_p$ available channels. We consider a single secondary user. The SU utilizes narrow-band spectrum sensing, i.e., it senses only one spectrum (out of $N_p$ spectrums) at a time. So, the maximum possible transmission opportunities obtained after each sensing phase is one. Moreover, it is assumed that the SU always has packets to transmit, i.e., the SU starts its transmission when an opportunity is found. The SU senses the channels in an order determined by its sensing sequence. If the sensed channel is busy, the SU reconfigures its sensing circuitry in order to sense the next spectrum indicated in its sensing sequence. This process is referred to as hand-over. It is assumed that it takes a constant time $\tau_{ho}$ for the SU to do a HO. $\tau_{ho}$ does not depend on the amount of frequency shift required by the reconfiguration. Fig. 1 illustrates these procedures. Furthermore, time-slots used by the SU are not necessarily synchronous with time-slots of the PUs. The state of channel $k$ that is used by $k$-th PU at time slot $t$ is denoted by $s_k(t)$

$$s_k(t) = \begin{cases} 1 & : \text{ if channel k is occupied , or } \mathcal{H}_1 \\ 0 & : \text{ if channel k is idle , or } \mathcal{H}_0 \end{cases} \tag{1}$$

where $\mathcal{H}_0$ and $\mathcal{H}_1$ denote the absence and presence hypotheses of the PU, respectively. Spectrum sensing can be formulated as a binary hypothesis testing problem [3],

$$
\begin{cases}
\mathcal{H}_0 \; : \; y(n) = z(n) \; : \quad \text{channel is idle} \\
\mathcal{H}_1 \; : \; y(n) = u(n) + z(n) \; : \quad \text{channel is occupied}
\end{cases}
\tag{2}
$$

where $z(n)$ denotes samples of zero mean complex-valued Gaussian noise with independent and identical distribution (i.i.d), $u(n)$ denotes the PUs signal which is independent of $z(n)$, and $y(n)$ is the $n$-th sample of the received signal.

Generally, there are various PU detection schemes such as match filtering, cyclostationary feature detection, waveform-based sensing, and energy detection (ED) [8], among which ED is the most prevalent because of its low complexity and ease of implementation. Further, it does not require any information about the PUs' signal attributes [9]. Therefore, we consider the ED method here, in which the energy of the received signal is computed during a sensing time $\tau$, and then the result is compared with a predefined threshold to make the decision [9]. Let $N$ denote the number of samples of the received signal, i.e., $N = \tau f_s$, where $\tau$ is the sensing time and $f_s$ is the sampling frequency. By defining $X$ as a decision metric for the ED scheme, we have,

$$
X = \sum_{n=1}^{N} |y(n)|^2
\tag{3}
$$

where $y(n)$ denotes the samples of the received signal. Let $\lambda$ denote the threshold of the ED decision rule. The decision criteria is defined as,

$$
\begin{cases}
X < \lambda \; \equiv \mathcal{H}_0 \\
X \geq \lambda \; \equiv \mathcal{H}_1
\end{cases}
\tag{4}
$$

If $N$ is large enough, $X$ can be described by a Gaussian distribution [3]. Assume that $P_{fa}$, $P_d$, and $P_d^{min}$ denote the false alarm probability, detection probability, and minimum allowable detection probability (i.e., we must have $P_d \geq P_d^{min}$), respectively. It is shown in [3] that

$$P_{fa} = Q\left[\left(\frac{\lambda}{\sigma_u^2} - 1\right)\sqrt{\tau f_s}\right], \tag{5}$$

$$P_d = Q\left[\left(\frac{\lambda}{\sigma_u^2} - 1 - \gamma\right)\sqrt{\frac{\tau f_s}{1 + 2\gamma}}\right] \tag{6}$$

where $\sigma_u^2$ is the received energy of the PU signal and $\sigma_z^2$ is the noise variance. The received signal to noise ratio due to the PU activity is $\gamma = \frac{\sigma_u^2}{\sigma_z^2}$.

For $P_d = P_d^{min}$, we have [10]

$$P_{fa} = Q\left[\beta + \gamma\sqrt{\tau f_s}\right] \tag{7}$$

where $\beta = Q^{-1}\left(P_d^{min}\right)\sqrt{1 + 2\gamma}$.

## B. Sensing Time Optimization Problem

In this paper, a sequential method for HO is used [5]. That is, the SU assigns a number to each of the channels and then arranges these frequency channels in a list by the assigned numbers. If HO is necessary, the SU starts sensing the spectrums from top of the list and continues this sequential sensing until a transmission opportunity is found.

The occupation probability of the $k$-th spectrum can be calculated as

$$\begin{aligned} q_k &= \Pr\{\text{ED says } 1|\mathcal{H}_0\}\, P_{k,0} + \Pr\{\text{ED says } 1|\mathcal{H}_1\}\, P_{k,1} \\ &= P_{fa}P_{k,0} + P_d P_{k,1}, \end{aligned} \tag{8}$$

where $P_{k,0}$ and $P_{k,1}$ are the absence and presence probabilities of the $k$-th PU, respectively. Let $\alpha$ denote the maximum number of allowable HO's. Clearly, $\alpha$ is limited by two constraints.

First, the number of sensed channels cannot exceed the number of the PUs. Second, the elapsed time for both sensing and HO procedures cannot exceed the time slot duration, $T$. So we have,

$$\alpha = \min\left(\left\lfloor \frac{T - \tau}{\tau + \tau_{ho}} \right\rfloor, N_P - 1\right) \tag{9}$$

The average throughput of the SU can be obtained as [10]

$$R = \sum_{m=0}^{\alpha} q_0 q_1 \cdots q_m \left(C_1 P_{m+1,1} \left(1 - P_d\right) + C_0 P_{m+1,0} \left(1 - P_{fa}\right)\right) \left(1 - \frac{\tau + m\left(\tau + \tau_{ho}\right)}{T}\right) \tag{10}$$

where $q_0 \overset{\Delta}{=} 1$, $C_0 = \log_2\left(1 + \gamma_s\right)$ and $C_1 = \log_2\left(1 + \frac{P_s}{N_0 + P_p}\right) = \log_2\left(1 + \frac{\gamma_s}{1 + \gamma_p}\right)$ are the SU's capacities under the hypotheses $\mathcal{H}_0$ and $\mathcal{H}_1$, respectively. $\gamma_s$ and $\gamma_p$ are the received SNRs due to the secondary and primary user's signals at the SU receiver, respectively. $P_{i,j}$ is as defined in (8).

The optimum throughput can be obtained by solving the following optimization problem $P1$:

$$P1: \quad \max_{\tau, \lambda} \quad R$$
$$s.t. \begin{cases} P_{fa} \leq P_{fa}^{\max} \\ P_d \geq P_d^{\min} \\ 0 < \tau < T \end{cases} \tag{11}$$

The derived optimization problem cannot be simplified to a one-dimensional one, as [3], without any preassumptions on $P_{fa}$ or $P_d$. However, we can convert our two-dimensional optimization problem to a one-dimensional one by using an acceptable value for the detection probability. Supposing $P_d = P_d^{\min}$, the optimization problem is converted to

$$P2: \max_{\tau, \lambda} \quad R$$

$$s.t. \begin{cases} P_{fa} \leq P_{fa}^{\max} \\ P_d = P_d^{\min} \\ 0 < \tau < T \end{cases} \tag{12}$$

It is worth noting that, under the assumption $P_d = P_d^{\min}$, the throughput of the SU derived in (10) only depends on $\tau$.

In order to satisfy the first constraint, from (7), we have $Q\left[\beta + \gamma\sqrt{\tau f_s}\right] \leqslant P_{fa}^{\max}$, So: $\tau \geqslant \frac{1}{f_s}\left(\frac{Q^{-1}\left(P_{fa}^{\max}\right) - \beta}{\gamma}\right)^2$. Therefore, $P2$ can be easily simplified as

$$P3: \max_{\tau} \quad R$$

$$s.t. \quad \tau_{\min} < \tau < T \tag{13}$$

where $\tau_{\min} = \frac{1}{f_s}\left(\frac{Q^{-1}\left(P_{fa}^{\max}\right) - \beta}{\gamma}\right)^2$.

Considering multipath fading, (10) is modified to

$$\overline{R} = \iint_{\gamma_p, \gamma_s} R \; f_{\gamma_p, \gamma_s}\left(\gamma_p, \gamma_s\right) d\gamma_p d\gamma_s \tag{14}$$

and then our optimization problem aims to maximize $\overline{R}$ instead of $R$ under the same constraint on $\tau$.

## C. Energy-Throughput Tradeoff

In the previous section, we formulated an optimization problem to maximize the average achievable throughput of the SU through choosing an appropriate value for the sensing time. In this section, we formulate the energy consumed for finding a transmission opportunity and we discuss the impact of the number of the PUs (i.e., $N_p$) on the SU throughput and consumed energy. Then, we propose a design parameter to control the trade-off between the consumed energy and achieved throughput.

Let $\overline{g}$ denote the average number of HOs required for finding a free transmission opportunity. Thus, $(1 + \overline{g})$ denotes the average number of sensed channels, and can be easily computed as,

$$1 + \overline{g} = 1 + q_1(1 - q_2) + 2q_1q_2(1 - q_3) + \cdots + (\alpha - 1)(1 - q_\alpha)\prod_{j=1}^{\alpha-1} q_j + \alpha \prod_{j=1}^{\alpha} q_j \qquad (15)$$

where $q_i$ is defined in (8) and $\alpha$ is the maximum number of allowable HOs which is defined in (9). Note that for the sequential handover method described in the previous sub-section, the probability of having $i$ consecutive handovers and transmitting on $(i + 1)$-th channel is equal to $(1 - q_{i+1})\prod_{k=1}^{i} q_k$. Thus, the average number of sensed channels can be determined by (15). Assume that $E_c(\tau)$ and $E_c(\tau_{ho})$ denote the consumed energy for sensing of each primary channel and the consumed energy during each HO, respectively. Hence, the average consumed energy for finding a transmission opportunity can be computed as,

$$E = (1 + \overline{g})E_c(\tau) + \overline{g}E_c(\tau_{ho}) \qquad (16)$$

Generally speaking, the processes of channel sensing and signal transmission consume more energy compared to the HO. Therefore, it is reasonable to ignore the second term $\overline{g}E_c(\tau_{ho})$ in (16) compared to the first one.

When the number of PUs increases, $\alpha$ in (9) increases and thus $(1 + \overline{g})$ and $\overline{R}$ in (15) and (14) increase, consequently. So increasing the number of PUs, increases both the number of sensed channels (so the consumed energy) and the maximum achievable throughput. But, among these two metrics, the consumed energy increases more rapidly than the maximum throughput. To illustrate this phenomenon, in the following we consider an numerical example.

Fig. 2 shows the plot of the normalized energy consumed for finding a transmission opportunity (i.e., $\frac{E}{E_c}$ which is equal to the number of sensed channels) versus the maximum achievable throughput assuming $P_{fa}^{\max} = 0.1$, $P_d^{\min} = 0.9$, $T = 100$ ms, $\tau_{ho} = 0.1$ ms, and $f_s = 6$ MHz. Simulation setup procedure is described in Section IV. As illustrated in Fig. 2, for high throughput values, increasing the consumed energy cannot improve the rate significantly. In fact, since increasing the number of HOs (i.e., increasing the consumed energy) reduces the time duration used for the transmission (for a fixed slot duration, $T$), the achievable rate is not substantially improved. For instance, for $N_p = 4$ with the average number of sensed channels equal to 1.8757, the maximum rate is 0.8544 while for $N_p = 15$ with the average sensed channels of 3.398, the maximum rate is 0.8809. As a result, increasing the average number of sensed channels by $81\%$, only leads to near $3\%$ increase in the maximum rate. Therefore, at the cost of small reduction of the maximum rate, the consumed energy can be substantially decreased. To take into account the energy consumption, in the following, we reformulate the optimization problem.

At first, we prove that $\overline{R}$ in (14) does not depend on the number of the PUs for sufficiently large $N_p$. If we rewrite $\overline{R}$, from (14) we have,

$$
\overline{R}\left(\tau, \lambda, N_P\right) = \iint_{\gamma_p, \gamma_s} \sum_{m=0}^{\alpha(\tau, N_P)} A_m\left(\tau, \lambda\right) B_m\left(\tau, \lambda\right) f_{\gamma_p, \gamma_s}\left(\gamma_p, \gamma_s\right) d\gamma_p d\gamma_s
$$

$$
\text{where} \begin{cases} A_m\left(\tau, \lambda\right) = C_1 P_{m+1,1}\left(1 - P_d\right) + C_0 P_{m+1,0}\left(1 - P_{fa}\right) \\ B_m\left(\tau, \lambda\right) = q_0 q_1 \cdots q_m \left(1 - \frac{\tau + m(\tau + \tau_{ho})}{T}\right) \\ \alpha\left(\tau, N_P\right) = \min\left(\left\lfloor \frac{T - \tau}{\tau + \tau_{ho}} \right\rfloor, N_P - 1\right) \end{cases} \tag{17}
$$

where $\overline{R}\left(\tau, \lambda, N_P\right)$ is the rate as a function of $\tau$, $\lambda$, and $N_p$. Considering the constraint of derived optimization problem imposed by sensing time, i.e, $\tau_{min} < \tau < T$, if $N_p \geq \left\lfloor \frac{T - \tau_{min}}{\tau_{min} + \tau_{ho}} \right\rfloor + 1$, we have,

$$\alpha\left(\tau, N_P\right) = \left\lfloor \frac{T - \tau}{\tau + \tau_{ho}} \right\rfloor = \alpha\left(\tau\right) \tag{18}$$

so

$$\overline{R}\left(\tau, \lambda, N_p\right)_{|N_p \geq \left\lfloor \frac{T - \tau_{min}}{\tau_{min} + \tau_{ho}} \right\rfloor + 1} = \overline{R}\left(\tau, \lambda\right) \tag{19}$$

If we indicate the maximum achievable throughput of the SU by $L$, from (19) we have,

$$L = \max_{\tau}\ \overline{R}\left(\tau, \lambda, N_P = \left\lfloor \frac{T - \tau_{min}}{\tau_{min} + \tau_{ho}} \right\rfloor + 1\right) = \overline{R}\left(\tau, \lambda\right)$$
$$s.t.\ \ \tau_{min} < \tau < T \tag{20}$$

To take into account the energy consumption, we define

$$\tau_{opt} = \arg\max_{\tau}\ \overline{R}\left(\tau, \lambda, N_P = \left\lfloor \frac{T - \tau_{min}}{\tau_{min} + \tau_{ho}} \right\rfloor + 1\right)$$
$$s.t.\ \ \tau_{min} < \tau < T \tag{21}$$

and the corresponding optimum number of HO is

$$\alpha_{opt} = \alpha_{|\tau = \tau_{opt}} \tag{22}$$

As stated above, at the rate close to the maximum achievable rate, i.e., $L$, defined in (20), by a small reduction of the rate, the average energy consumption is substantially reduced. Now, let us define $TF(0 \leq TF \leq 1)$ as a Tradeoff Factor indicating the amount of the rate reduction considered. That is the target rate is set as $R_{TF} = \text{TF} \times \text{L}$. Then, from (20), the maximum number of HOs $\overline{\alpha}$ ($\overline{\alpha} \leq \alpha_{opt}$) considering the energy consumption concern (reflected in the parameter $TF$) is obtained by solving the following equation:

$$TF = \frac{\iint\limits_{\gamma_p, \gamma_s} \sum\limits_{m=0}^{\overline{\alpha}} A_m(\tau, \lambda) B_m(\tau, \lambda) f_{\gamma_p, \gamma_s}(\gamma_p, \gamma_s) d\gamma_p d\gamma_s}{L} \qquad (23)$$

$\overline{\alpha}$, the maximum number of SU's HO, is selected as follows. From (20), $L$ is calculated and by choosing a value for $TF$, $\overline{\alpha}$ is obtained from (23). Finally a new optimization problem considering the consumed energy is formulated as:

$$\max_{\tau} \ R_{TF} = \iint\limits_{\gamma_p, \gamma_s} \sum\limits_{m=0}^{\bar{\alpha}} \left( \begin{array}{l} (C_1 P_{m+1,1}(1-P_d) + C_0 P_{m+1,0}(1-P_{fa})) \times \\[2mm] q_0 q_1 \cdots q_m \left( 1 - \dfrac{\tau + m(\tau + \tau_{ho})}{T} \right) \end{array} \right) f_{\gamma_p, \gamma_s}(\gamma_p, \gamma_s) d\gamma_p d\gamma_s$$

$$s.t. \ \ \tau_{\min} < \tau < T$$

$$\qquad (24)$$

This new derived optimization problem enables the SU to have control on the consumed energy and the achieved average throughput by the parameter $TF$.

## III. Neural Network-Based Optimization Scheme

### A. Learning and Optimization

Artificial neural networks are powerful tools for learning complicated mappings and for optimization. Feed forward neural networks have been shown to be capable of approximating any arbitrary nonlinear mapping and its partial derivatives. Specifically, Hornik *et al.* [11], [12] showed that multilayer feed forward neural networks with as few as a single hidden layer and an appropriately smooth hidden layer activation function are capable of providing arbitrarily accurate approximation of almost any given function and its derivatives. Morover, the Kennedy-Chua neural networks that are a special type of the classical Hopfield network [13] can solve general nonlinear programming problems. As described in [14], the Kennedy-Chua neural networks are widely known in the literature for their capabilities in minimizing cost functions, and among other advantages, they can be entirely made of simple electronic devices such as capacitors, resistors, and operational amplifiers, and also suitable for the implementation in a very large scale integration (VLSI) technology.

The optimization problem can be divided into two distinct parts. First, obtaining the mappings between $\tau$ and $\overline{R}$ that is described by (14). Second, from (13) finding the optimum value for $\tau$ using the derived mapping. In the proposed learning-based optimization scheme described in Section III, both of these parts are performed effectively using artificial neural networks.

We have to restate the spectrum sensing time optimization problem suitable for the proposed optimization scheme. So we define our adaptable parameter $x$, the cost function $\varphi$, and the constraint functions $f_1$ and $f_2$ as

$$x \triangleq \tau$$

$$\varphi\left(x\right) \triangleq 1/\bar{R}\left(x\right)$$

$$f_1\left(x\right) \triangleq T - x \tag{25}$$

$$f_2\left(x\right) \triangleq x - \tau_{\min}$$

Hence, from (13) and (25), the optimization problem can be rewritten as

$$x^{opt} = \underset{x}{\arg\min} \quad \varphi(x) \tag{26}$$

$$\text{s.t.} \quad f_1(x) \geq 0 \tag{27}$$

$$f_2(x) \geq 0 \tag{28}$$

We collect the constraint functions and their derivatives in a matrix named $\mathbf{F}$ and defined as

$$\mathbf{F} = \begin{bmatrix} f_1(x) & f_2(x) \\ f_1{}'(x) & f_2{}'(x) \end{bmatrix} \tag{29}$$

The MFF network has one input corresponding to $x$ and one output corresponding to the learned version of $\varphi(x)$. It also has one extra output corresponding to the sensitivity of the cost function, i.e., $\partial\varphi/\partial x$, see [15].

The exploited MFF network consists of 1-K-1 neurons at the input layer, the hidden layer, and the output layer, respectively. The output of the $i$-th neuron at the $l$-th layer is described as,

$$u_i(l) = \sum_{j=1}^{N_l-1} w_{ij}(l)a_j(l-1) + b_i(l) \tag{30}$$

$$a_i(l) = h_l(u_i(l)), \quad 1 \leq i \leq N_l, \quad l = 1, 2 \tag{31}$$

where $N_l$ is the number of neurons at the $l$-th layer; and $u_i(l)$ and $a_i(l)$ are the activation and output values of the $i$th neuron at the $l$-th layer. $w_{ij}(l)$ refers to the weight connecting the output from the $j$-th neuron at the $(l-1)$-th layer to the input of the $i$th neuron at the $l$-th layer. $b_i(l)$ refers to the bias associated with the $i$-th neuron at the $l$-th layer. The utilized transferring function $h_l(\cdot)$ in (31) is logistic sigmoid at hidden layer ($l = 1$) and is hyperbolic tangent sigmoid at output layer ($l = 2$), i.e.,

$$h_l(x) = \begin{cases} (1 + e^{-x})^{-1}, & l = 1 \\ 2(1 + e^{-2x})^{-1} - 1, & l = 2 \end{cases} \tag{32}$$

The input unit is demonstrated by $a_i(0)$ and the output unit by $a_i(2)$, so we have,

$$x = a_1(0)$$
$$\hat{\varphi}(x) \overset{\Delta}{=} a_1(2) \tag{33}$$

where $\hat{\varphi}$ is the learned version of the cost function. We denote the set of weights and biases by the matrix $\mathbf{w}$.

The MFF network can be trained to model the function $\varphi$, by recursively adjusting $w_{ij}(l)$ and $b_i(l)$ to minimize the mean-squared error (MSE) between the MFF network output $\hat{\varphi}$ and our cost function $\varphi$,

$$d = \frac{1}{2} \sum_{m=1}^{M} |\hat{\varphi}_m - \varphi_m|^2 \tag{34}$$

where $M$ is the number of teacher patterns.

The first order output derivative of the MMF networks, i.e., $\partial a_1(2)/\partial a_1(0)$, can be calculated by applying a backward chaining partial differentiation rule that is described in detail in [15]:

$$\frac{\partial a_j(2)}{\partial a_i(0)}$$

$$= \sum_{k=1}^{N_1} \frac{\partial a_j(2)}{\partial a_k(1)} \cdot a_k(1) \left[1 - a_k(1)\right] \cdot w_{ki}(1)$$

$$= \sum_{k=1}^{N_1} \frac{\partial a_j(2)}{\partial u_j(2)} \cdot \frac{\partial u_j(2)}{\partial a_k(1)} \cdot a_k(1) \left[1 - a_k(1)\right] \cdot w_{ki}(1)$$

$$= \left[1 - a_j(2)^2\right] \cdot \sum_{k=1}^{N_1} w_{jk}(2) \cdot a_k(1) \left[1 - a_k(1)\right] \cdot w_{ki}(1) \qquad (35)$$

The KC network uses the MFF network output, $\hat{\varphi}(x)$, and its derivative to solve the optimization problem.

The KC neural network has one output voltage corresponding to the adaptable parameter $x$. This network calculates the optimum sensing time based on the cost function learned by the MFF network. Now, if there exist a training process to adjust the weight and bias values of the MFF network appropriately and if the learned mapping approximates the actual cost function closely, then the KC network output is equal to the optimum value of $x$ (i.e., the sensing time).

The dynamic equation implemented by the KC neural network is [7], [16]:

$$C\frac{dx}{dt} = -\frac{\partial \hat{\varphi}}{\partial x} - \sum_{j=1}^{2} i_j \frac{\partial f_j}{\partial x} - Gx \qquad (36)$$

where $i_j = g_j\left(f_j\left(x\right)\right)$, and $C$ and $G$ are the output capacitor and the parasitic conductance of the KC network, respectively, and $g_j\left(.\right)$ is defined as [7]

$$g_j(v) = \begin{cases} 0, & v \geq 0 \\ \frac{1}{R}v, & v \leq 0 \end{cases}, \quad R \to 0 \qquad (37)$$

In the next sub-section, a detailed description of the proposed adaptive system is presented.

*B. Proposed Scheme*

Fig. 3 demonstrates the proposed neural network-based optimization scheme. It consists of a KC neural network cooperating with an MFF neural network in a feedback loop, a training process which calculates and updates the weight and bias values of the MFF network, and a throughput estimator (TE).

The TE estimates the SU throughput and calculates the value of $\varphi(x)$. This estimation can be performed by inspecting the packets and their acknowledgments (ACKs) at the secondary transmitter for a period of time equal to $T_{ep}$ (estimation period) [17]. $T_{ep}$, as a design parameter, depends on the PUs activity and link behavior.

As mentioned before, the KC neural network has one output corresponding to $x$. It calculates the optimum value for $\tau$ based on the cost function provided by the MFF network $\hat{\varphi}(x)$. Its output, even though not necessarily optimal at first, is always used by the ED as the channel sensing time. The learned mapping of the MFF network is considered as the cost function by the KC network. Specifically, this learned function, i.e., $\hat{\varphi}(x)$, and its derivative are used by the KC network to establish its dynamic equation (36). Thus, the KC network output $x$ will be sufficiently close to the optimum value provided that the function learned by the MFF network can model the link behavior sufficiently accurate, i.e., $\hat{\varphi}(x) \approx \varphi(x)$. Once the KC network output $x$ is applied to set the spectrum sensing time of the ED, the TE estimates the throughput obtained by this setting and calculates $\varphi(x)$ within $T_{ep}$ seconds. Then, the training process uses $x$ as the input and the estimated $\varphi(x)$ as the target to adjust the weight and bias values of the MFF network by the well-known backpropagation algorithm [6]. Having its weight and bias values modified, the MFF network models the link more accurately, and therefore the KC network output takes a new value closer to the optimal point $x^{opt}$. By iterating this learning and optimization cycle, we observe a joint convergence in the weight and bias values $\mathbf{w}$ and more importantly in the KC network output $x$. That is, $\mathbf{w}$ converges to $\mathbf{w}^{opt}$ by which the learned mapping fits the cost function $\varphi(x)$ appropriately and $x$ converges to $x^{opt}$ which denotes the optimum value for $\tau$, thanks to universal approximation theorem [12].

To sum up, the proposed adaptive system works according to the following three-steps algorithm:

Step 1) *Setting*: The KC network output $x$ is applied to the ED to set its sensing time duration.

Step 2) *Throughput Estimation*: The average throughput is estimated by inspecting the packets and their ACKs for $T_{ep}$ seconds, and accordingly $\varphi(x)$ is calculated.

Step 3) *Training*: The KC network output $x$ and the TE outputs $\varphi(x)$ are used as an input-target pair to adjust the weights and biases of the MFF network, and then the process returns back to *Step 1)*.

Computational complexity of the proposed system is due to the backpropagation algorithm. The complexity order of the backpropagation is $\mathcal{O}(N)$, where $N$ is the number of weights and biases of the MFF network [6]. For the examples considered in numerical result section, 9 hidden neurons have been enough to well model the relationship between sensing time and SU's throuhput. The MFF network in the proposed system with 9 hidden neurons can be considered equivalently as an adaptive filter with $N = 1 \times 9 + 1 \times 9 + 10 = 28$ weights which are being updated by the LMS algorithm.

## IV. NUMERICAL RESULTS

In this section, we first evaluate the maximum throughput achieved by the SU when SMHO scheme is used based on both the analytical results and simulations. The impact of various sensing time durations as well as various number of primary users are considered in this evaluation. Then, the performance of the proposed learning-based scheme is demonstrated through simulations.

Our numerical parameters are described in Table I. The values of SNR and sampling frequency are adopted from [3], and $P_d^{\min}$ and $P_{fa}^{\max}$ are chosen according to *IEEE 802.22* standard [18]. In simulation evaluations, the average throughput has been computed after simulating the scenario for 100 time slots.

Fig. 4 verifies our analysis and depicts the achievable rate versus normalized sensing time (i.e., $\frac{\tau}{T}$) for various values of $N_p$ (number of primary users) assuming that the presence probabilities of all the PUs are equal to 0.65. Fig. 4 shows that, for large normalized sensing time, the plots for different values of $N_p$ coincide.

This behavior is expected due to our previous discussions on the constraints which affect the number of possible HOs for a SU. As stated previously in Section II, the number of possible HOs is dictated by two factors; namely, the number of primary channels $N_p$, and the ratio $(T - \tau) / (\tau + \tau_{ho})$. Therefore, as $\tau$ increases in Fig. 4, we observe that the second factor dominates and regardless of the number of available primary channels $N_p$, the achieved throughput becomes limited to a value corresponding to a lower $N_p$. The observed coincidences of the plots in Fig. 4 demonstrate this effect. The rate of the SU where there are 10 primary channels equals the rate of the SU with 3 primary channels for approximately $\tau > 1/4T$. Similarly, the rate achieved by a SU with 3 primary channels is equal to the rate of a SU with only 1 primary channel. Other important observations can be made through Fig. 4. First, since all the curves posses a maxima, there exists an optimum value for the spectrum sensing time. Second, as the number of primary channels increases, the SU throughput increases as well, but in a saturating manner. This is due to the fact that, as the number of primary channels increases, the average number of obtained transmission opportunities increases as well, but the average time duration in which the SU transmits data reduces.

The effectiveness of the discussed energy-throughput tradeoff on the proposed sensing time optimization scheme is demonstrated by Table II. In this table, two design sets are presented namely *Design 1* and *Design 2*. *Design 1* is referred to the analytical optimization that does not consider the energy consumed by the SU, whereas in *Design 2*, the sensing time is optimized using the $TF$ factor to obtain an energy-efficient sensing scheme. This table illustrates that considering energy-throughput tradeoff provides near maximum throughput for each $N_p$ with much lower consumed energy compared to *Design 1*. It is worth noting that the SU consumed energy can be very high when $N_p$ is large. Therefore, Table II shows that the proposed energy-efficient sensing time optimization procedure reduces the consumed energy dramatically when the consumed energy concern is more serious, i.e., when the number of primary channels is large. See the $65\%$ reduction in consumed energy for *Design 2* compared to *Design 1* at the expense of only $2\%$ reduction in the throughput when $N_p = 12$ for the example considered.

Figs. 5 and 6 demonstrate the performance of the proposed learning-based optimization scheme. They compare the maximum normalized throughput and the optimum sensing time obtained by our learning-based optimization scheme with the results achieved by the analytical modeling. This comparison is performed for various number of primary channels (i.e., various $N_p$s), for the presence probabilities of the PUs given in Table III. Results labeled as *Analysis* depict the maximum normalized throughput and optimum sensing time obtained by the analytical modeling and results labeled as *Learning-based* are the ones obtained through the neural network-based optimization. As can be realized from Fig. 5, the average throughput values obtained by the proposed neural network-based optimization method are very close to the maximum possible values obtained through the the mathematical analysis. Moreover, Fig. 6 shows that optimum sensing times calculated through our learning-based optimization method are very close to the results obtained by the analysis.

Fig. 7 depicts the KC network output convergence to the optimal sensing time corresponding to a secondary link with $N_p = 3$ primary channels. As shown in this figure, using the learned cost function provided by the MFF network, the KC output converges to the optimal sensing time very fast. Therefore, we observe that the KC neural network locks its output to the minimum point of the cost function learned by the MFF network and if the learned function well-approximates the actual link model, the KC output represents the optimal value of the sensing time.

## V. Conclusion

In this paper, we have considered the problem of channel sensing time optimization for a secondary user which senses the primary channels sequentially. Maximizing the average throughput of the secondary user by optimizing the spectrum sensing time has been formulated assuming that a prior knowledge of the presence and absence probabilities of the primary users are available, and then the energy-throughput tradeoff of a CR system has been discussed using the derived expressions. Moreover, a learning-based sensing time optimization approach has been proposed using a novel and effective combination of two powerful and well-organized artificial neural networks and a complete description of its structure and operational algorithm have been presented. Finally, the validity of our analytical results as well as the capability of the proposed adaptive system in finding the optimal spectrum sensing time have been demonstrated by a set of illustrative simulation results.

TABLE I

SIMULATION PARAMETERS

| Parameter | Description | Value |
|:---:|:---:|:---:|
| $P_d^{\min}$ | Minimum allowable detection probability | 0.9 |
| $P_{fa}^{\max}$ | Maximum allowable false alarm probability | 0.1 |
| $f_s$ | Receiver sampling frequency | 6 MHz |
| $T$ | Time-slot duration | 100 ms |
| $\tau_{ho}$ | Required time for handover | 0.1 ms |
| $N_p$ | Number of primary users | 15 |
| $K$ | Number of hidden neurons | 9 |
| $C$ | Capacitances of Kennedy-Chua network | 10nF |
| $G$ | Conductances of Kennedy-Chua network | $0.001\Omega^{-1}$ |

TABLE II

THE SU'S AVERAGE THROUGHPUT AND NORMALIZED CONSUMED ENERGY (NCE) FOR $TF = 1$ IN *Design 1* (WITHOUT CONSIDERING ENERGY-THROUGHPUT TRADEOFF) AND $TF = 0.98$ IN *Design 2*

| | **Design 1** | | **Design 2** | |
|:---:|:---:|:---:|:---:|:---:|
| | Throughput | NCE | Throughput | NCE |
| $N_p = 2$ | 0.775 | 1.3186 | 0.775 | 1.3186 |
| $N_p = 6$ | 0.8723 | 2.3909 | 0.8660 | 2.1397 |
| $N_p = 12$ | 0.8807 | 3.3080 | 0.8660 | 2.1397 |

TABLE III

PRIMARY FREE PROBABILITIES

| $k$ | 1 | 2 | 3 | 4 | 5 | 6 | 7 | 8 | 9 | 10 | 11 | 12 | 13 | 14 | 15 |
|---|---|---|---|---|---|---|---|---|---|---|---|---|---|---|---|
| $P_{k,0}$ | 0.71 | 0.46 | 0.34 | 0.72 | 0.66 | 0.72 | 0.76 | 0.35 | 0.25 | 0.70 | 0.37 | 0.23 | 0.72 | 0.24 | 0.43 |

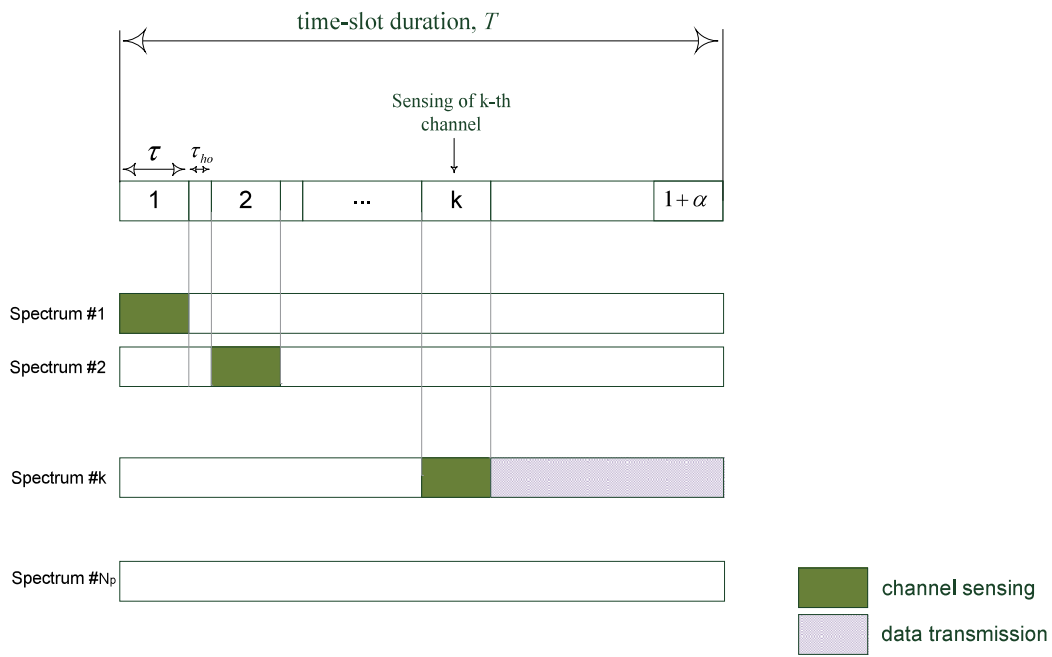

Fig. 1.   General structure of time-slots in our system model .

Fig. 2.   The relationship between the normalized consumed energy of a SU and its maximum achievable throughput.

Fig. 3.   Block diagram of the proposed neural network-based sensing time optimization scheme.

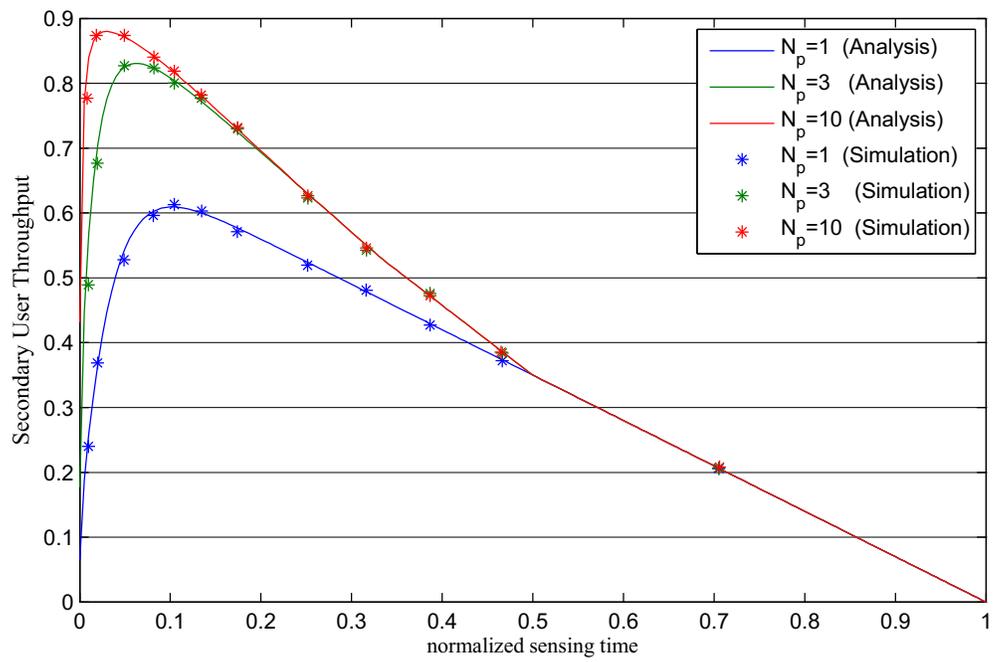

Fig. 4.   Secondary user throughput versus normalized sensing time.

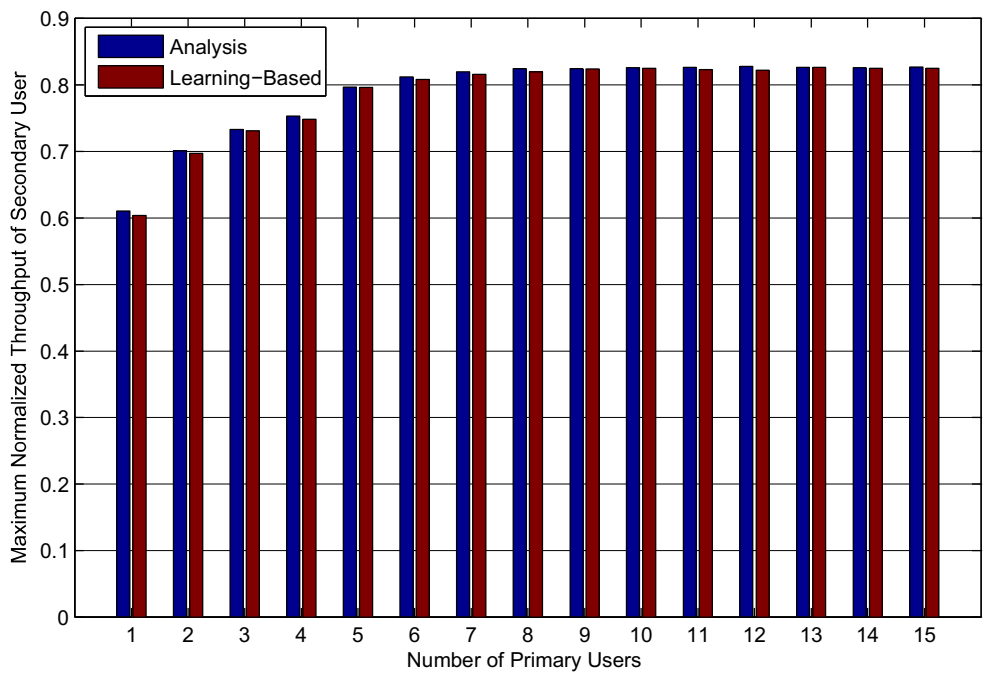

Fig. 5.   Maximum normalized throughput of the secondary user versus the number of channels.

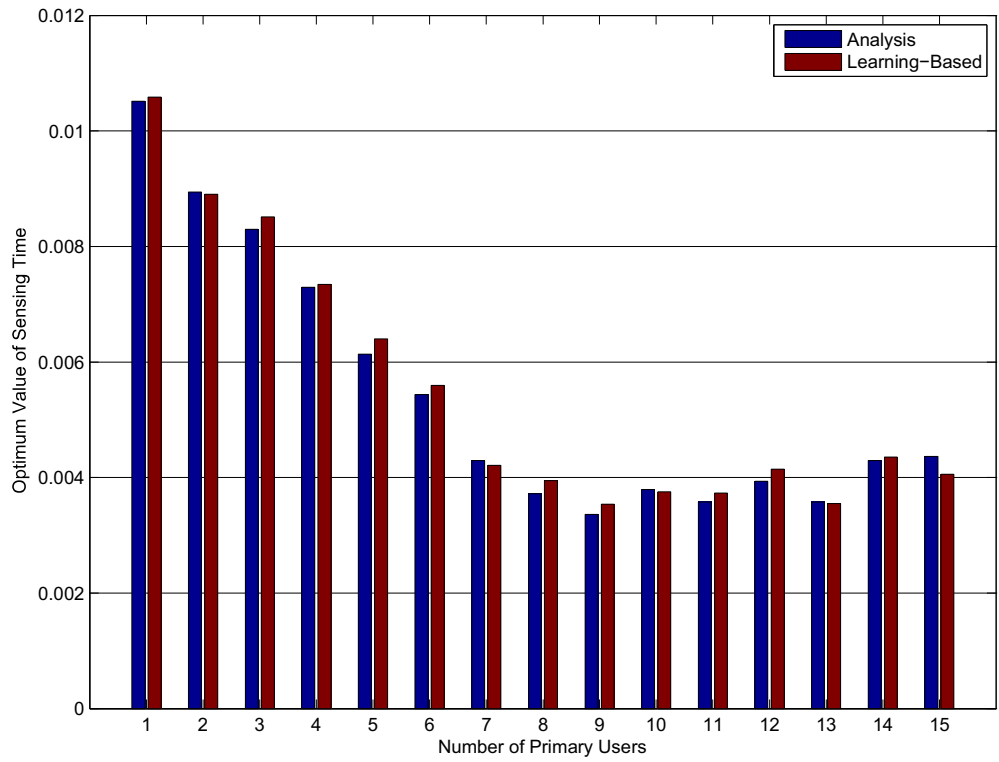

Fig. 6. Optimum value of the sensing time obtained by the two methods.

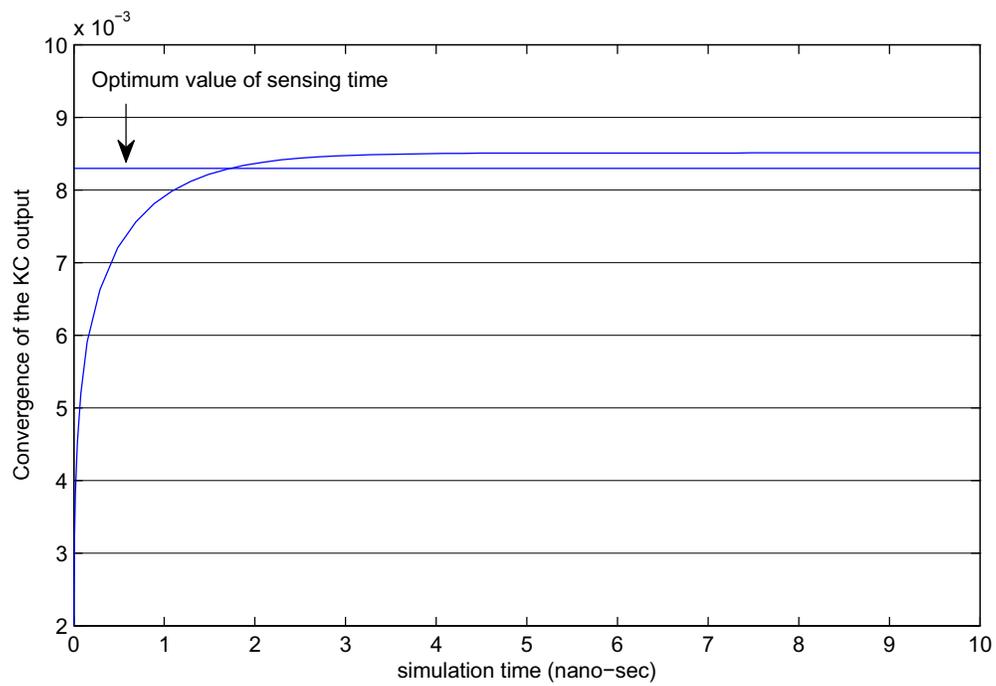

Fig. 7.   Convergence of the output of the KC neural network to optimal sensing time using a learned model for a secondary

link with $N_p = 3$ primary channels.